\documentclass[12pt,preprint]{aastex}

\def\gs{\mathrel{\raise0.35ex\hbox{$\scriptstyle >$}\kern-0.6em
\lower0.40ex\hbox{{$\scriptstyle \sim$}}}}
\def\ls{\mathrel{\raise0.35ex\hbox{$\scriptstyle <$}\kern-0.6em
\lower0.40ex\hbox{{$\scriptstyle \sim$}}}}

\shortauthors{Owen et al}
\shorttitle{Deep SWIRE Field II: 90cm Continuum}

\begin{document}

\title{The Deep SWIRE Field}
\title{II. 90cm Continuum Observations and 20cm--90cm Spectra}

\author{
Frazer N. Owen,\altaffilmark{1}, 
G.\,E.\ Morrison,\altaffilmark{2,3}
 Matthew D. Klimek,\altaffilmark{1,4}
\& Eric W. Greisen, \altaffilmark{1}}
\altaffiltext{1}{National Radio Astronomy Observatory, P.\ O.\ Box O,
Socorro, NM 87801 USA.; The National Radio Astronomy
Observatory is facility of the National Science Foundation operated
under cooperative agreement by Associated Universities Inc.}
\altaffiltext{2}{Institute for Astronomy, University of Hawaii,
  Honolulu, Hawaii, 96822,
  USA}
\altaffiltext{3}{Canada-France-Hawaii Telescope, Kamuela, Hawaii,
  96743, USA}
\altaffiltext{4}{Department of Physics \& Astronomy, Rutgers University,
136 Frelinghuysen Rd, Piscataway, NJ 08854, USA}

\setcounter{footnote}{4}

\begin{abstract}

We present one of the deepest radio continuum surveys to date at a wavelength
$\gtrsim 1$ meter.  The observations were taken with the VLA at 324.5
MHz covering a region of the SWIRE Spitzer Legacy survey, centered
at 10$^h$46$^m$00$^s$, 59\arcdeg01\arcmin00\arcsec\ (J2000).  The
data reduction and analysis are described and an
electronic catalog of the sources detected above 5 sigma is presented.
We also discuss the observed angular size distribution for the sample.

Using our deeper 20cm survey of the same field, we calculate
spectral indices for  sources detected in both surveys.
The spectral indices for 90cm-selected sources, defined as
$S \propto \nu^{-\alpha}$, shows a peak near 0.7
and only a few sources with very steep spectra, i.e
$\alpha^{90}_{20} >> 1$. Thus no large population of very steep
spectrum $\mu$Jy sources seems to exist down to the
limit of our survey.

For 20cm-selected sources, we find similar mean spectral indices
for sources with $S_{20}>1$ mJy.
For weaker sources, below the detection limit for individual sources
at 90cm, we use stacking to study the radio spectra.  We find
that the spectral indices of small ($<3$\arcsec) 20cm-selected sources
with $S_{20}< 10$ mJy have mean and median $\alpha^{90}_{20} \sim
0.3-0.5$.  This is flatter
than the spectral indices of the stronger source population.
At the low end of the 20cm survey, the spectral indices appear to be
steepening again. 

We  report log N -- log S counts at 90cm which show a  flattening  below
5 mJy.  Given the median redshift of the population,
$z\sim 1$, the spectral flattening and the flattening of the 
log N -- log S counts occurs at radio luminosities normally associated with
AGN rather than with galaxies dominated by star-formation. 

\end{abstract}


\section{Introduction}

We are building a deep multi-wavelength picture of the sky in
the SWIRE Spitzer deep field, 1046+59, which was chosen to be ideal 
for deep radio imaging. In paper I we discussed the 20cm 
continuum survey . The present 90cm survey 
allows us to study
the radio spectra of the general source population. For Jansky
and mJy sources, very steep radio spectra often are associated with
very high redshifts, although the physical origin of this
effect remains unclear \citep[e.g.,][]{mi08}. A large
population of very steep spectrum, $\mu$Jy sources might suggest 
a corresponding high redshift $\mu$Jy population. On the
other hand, flatter radio spectra are often thought to be connected with 
synchrotron self-absorption or free-free absorption, although other mechanisms
could potentially produce such spectra.  Combined with other information
the low frequency spectral energy distribution has the potential to
give us unique insight on the physics of
black-hole-driven AGN and star-forming galaxies.
In this paper we report our 90cm observations with the VLA and some 
analysis of these radio data combined with our 20cm survey of the same
field from paper I. In future papers in this series, we will
combine these data with redshift measurements and observations at
other wavelengths.

\section{Observations, Reduction and Cataloging}

	Observations were made of a single pointing center position,
10$^h$46$^m$00$^s$, 59\arcdeg01\arcmin00\arcsec\ (J2000), with the VLA in A and C 
configurations for a total of almost 85 hours on-source between
February 2006 and January 2007. However, due to the ongoing EVLA upgrade, only
22 working antennas were typically available in A and 18 in C. Thus
the total integration time was equivalent to $\sim$63 hours in A and
$\sim$5 in C, with correspondingly less $uv$ coverage.  In
Table~\ref{OR}, 
we summarize the parameters of the observing runs.
Since the total time is dominated by the A configuration, the final image
for analysis had a resolution $\sim 6$\arcsec and FWHM FOV of 2.3\arcdeg.  The data
were all taken in spectral-line mode 4 using on-line Hanning smoothing, 
resulting in fifteen
390.625 kHz channels in each of 2 IFs (centered at 321.5 and
327.5 MHz) and each of two polarizations. Five second
integration times were used in the A configuration and 10 seconds in
C.  The integration
times and channel bandwidths were chosen to minimize 
tangential and radial smearing of the images away from the
field center. This combination of parameters produces the best
compromise for imaging sensitivity and quality possible with
the current VLA correlator, which dates from the 1970's.
The finite bandwidth of the spectral channels still produces some radial
smearing of the image away from the field center which we 
take into account in the analysis of the image.

\subsection{Calibration \& Editing and Imaging}

For calibration, editing, and imaging a procedure similar to the one
described in paper I was used. The Baars flux density
scale \citep{baars} was adopted using 3C286 as the flux
calibrator. 
Two of the 15 channels in each IF were deleted due to interference
which is generated by the VLA itself and which should disappear when the EVLA
is completed. Unless otherwise stated, the AIPS package \citep{g03} was used to
reduce these data. 

A faceted, low resolution image (90\arcsec\ clean beam) with a radius
of 15 degrees was made to find interfering sources far from the area
of interest. Facets centered on all very bright NVSS sources ($> 30$
Jy) out to 100 degrees
from the field center were also included in this exploratory image.
From this search 288 facets, each with $500\times 500$ pixels, were
chosen to cover a central region
93\arcmin\ in radius and all the other bright sources found in the low
resolution search. The facets were defined using the task SETFC which
creates a set of overlapping circular regions within the square facets
to cover the entire
desired field.  Then IMAGR was used to deconvolve all the facets
together, using the standard Cotton-Schwab-Clark clean 
algorithm \citep{s84}. 
The cell size for the final image is 2\arcsec\ and the clean beam size
is $6.37$\arcsec$\times5.90$\arcsec\ pa$=86\arcdeg$. 

Clean images from the first day of the observations 
were then used as fiducial models for each of the
other days. Phase and amplitude calibrations were made of each of the
other days using the clean components from the first day images. 
The A configuration data for each IF and polarization were
then combined into a smaller, averaged dataset using STUFFR and images
for the full datasets were made. 
The C configuration data were also calibrated using the full A
configuration images. The A and C datasets were then combined using
DBCON and images were made separately for each IF and polarization.

After making these images there remained some significant residual
structures in the central two degrees of the image due to bright
sources located outside the central region.  These
residuals are likely due to 1) differences in the primary beam
patterns from antenna to antenna due to the very simple dipole feeds
used on the VLA and 2) the rotation with parallactic angle of the
sensitivity pattern on the sky during the synthesis.  For bright
sources in the outskirts of the field, these variations in
sensitivity produce local gain variations which are not taken into
account in the imaging and self-calibration process.  In order to deal
with this type of error, an AIPS procedure, PEELR, was developed and
made generally available in the AIPS package. In PEELR the best clean
model is subtracted from the self-calibrated {\it uv} data, except for
the facet containing the bright source responsible for the
residuals.  These mostly residual data are then self-calibrated as
a function of time using only the model for the bright source.
This process allows the variations in gain due to the actual primary
beam of each antenna to be tracked in time.  This local complex
calibration is then applied to the mostly residual data and
the model for the bright source subtracted from the locally calibrated
result.  This operation removes, as accurately as possible, the
contribution of the offending source.  The inverse of the local
calibration is applied to the resulting fully residual data and
the full clean model added back to the {\it uv} database.  If
there are several offending sources, this process can be repeated for
each source as it was in the present case.  When the ``peeled''
dataset is then imaged, the effects of the offending outlying sources
are significantly reduced. Techniques like this one have been used by
others, but only by combining several different steps and perhaps not
quite in the same way as described here. 

The resulting images still showed radial
smearing for bright sources in excess of what is expected from the 
finite bandwidths. To
explore these errors we subtracted the clean component model from each
visibility dataset and made spectral line image cubes from the 
residual visibilities. For facets containing bright
sources far from the field center, these images showed
frequency-dependent artifacts which
are likely due to the different slopes of the bandpass across each
spectral channel. This instrumental problem causes the effective 
observing frequency for each channel to be 
slightly different than is assumed and thus the $uvw$
coordinate for each channel used in the imaging to be slightly in error. 
As a consequence, the source image in each channel is slightly
mis-registered,
producing a radial smearing for very bright sources far from the field
center. To remove the error pattern due to this effect we cleaned the
spectral residual image cubes in facets containing bright, outlying sources and
subtracted the resulting clean components from the corresponding
visibility data. This reduced the error pattern significantly. 

The resulting four sets of 288 facets were
then made into single images using FLATN. Finally a weighted average
of the four images was made, weighting by the 1/rms$^2$, as determined
from the IMEAN fit to the pixel histogram.  This final
image still showed a weak, large-scale error pattern due to the
imperfections of the corrections described above. For the final image used
for most of the analysis, the AIPS program MWFLT was used to calculate
the ``mode'' of the image over an $82\arcsec\times82\arcsec$ support window and the
result was subtracted from the image. For sources approaching this
scale size the image before MWFLT was used for analysis but for the
vast majority of the sources the MWFLT image was used. 

The rms noise near the center of the final image is $\sim 70\mu$Jy
beam$^{-1}$ making it one of the most sensitive surveys to date made at such a long
wavelength. The corresponding value for the
20cm image discussed in paper I is  much lower, $\sim 2.7\mu$Jy beam$^{-1}$. 
For a spectral index, $\alpha^{90}_{20} \sim 0.7$ the ratio of flux
densities between 90cm and 20cm is $\sim 2.8$. Thus the effective 
sensitivity difference in the field center is about one order of
magnitude. However the resolution for the full sensitivity image
is about $4\times$ worse at 90cm than at 20cm. Also the primary
beam at 90cm is about $4\times$ wider. Since, many of the sources
are resolved at 20cm and because most are more than a few arcminutes from the 
field center, the sensitivity difference is not as large as at the
field center and can be
better at 90cm than at 20cm  very far from the field center. Thus
in comparing the two surveys we need to keep in mind the local
properties of each source in the image of interest at each wavelength. 

\subsection{Cataloging}

Although the primary beam has a diameter of 2.3\arcdeg\ and
the region we imaged extends out far beyond this limit, we
chose to catalog the region within 1\arcdeg\ of the field
center.  This two-degree diameter field-of-view covers more than
the entire field we are studying at other wavelengths.  Moreover,
the primary beam shape becomes less well known beyond one degree and
the smearing due to the finite channel width begins to become
important beyond this radius.  As with the 20cm survey\, we
include the radial smearing due to the finite bandwidth in our
Gaussian fits.  In the present case the angular FWHM of the radial
smearing function is 
approximately  $0.0012\times$ (the distance from the field center).
 In the worst case, for a point source at the maximum
cataloged radius from the field center, this bandwidth smearing
amounts to a decrease in the peak brightness at full resolution of $\sim
18$\%.  As for the 20cm survey, we convolved our $\sim 6$\arcsec\
images to resolutions of 12\arcsec\ and 24\arcsec\ to increase the
detection sensitivity for large sources.  This exercise was much less
important than it was for the 20cm survey, but it did yield
higher S/N detections for a subset of the survey.  

	As for the 20cm survey, the AIPS program SAD was used for
forming the initial source lists. A
catalog for each resolution was formed down to a peak signal/noise (S/N) of
4.5. The residual images from SAD were then searched to find any
remaining sources  missed by the program with a S/N greater than
5.0. For sources with a S/N close to 5.0, the fitting process was repeated
by hand with JMFIT, using the local rms estimate over a
region 100 pixels in diameter. In this way a reliable list of sources with
S/N of 5.0 or greater at each resolution was compiled. For both SAD
and JMFIT, the smearing due to the finite bandwidth was included in
the fitting process. As described for the 20cm survey in paper I, 
the best description of each of the sources with a
peak S/N$\ge 5$ at one of the resolutions analyzed was included in the 
final 90cm catalog.  In order to allow for a
variety of potential calibration errors, a 3\% error term
proportional to the total measured flux density of each source was
folded into the errors in quadrature, as was done for the 20cm
survey.  

\section{Results}

\subsection{Radio Catalog}

	In Table~\ref{S90}, we give the first ten lines of the
radio catalog; the full table is provided electronically.  Column (1) 
contains the source number. If the source has
a number less than 3000, then it was found
with a S/N $\ge 5.0$ from running SAD on the full resolution image.
Sources with numbers $\ge 10000$ were found in lower resolution images
or in checks of the residual images. Numbers beginning with 12 were
measured on a 12\arcsec\ resolution image. If the number begins with
24 then we used a 24\arcsec\ resolution image. The lower resolution 
fits were used in the table when they indicated a significantly larger 
total flux density for the source in question. Each of these cases was
also investigated by eye to confirm the result. 
Columns (2) and (3) contain the radio RA and Dec along with the
estimated error. Column (4) contains the
 corrected peak flux density from the map in $\mu$Jy per
beam. In column (5) we list the corrected total flux density. In
column (6) we give the estimated error in the total flux density. 
Column (7) contains the peak S/N. The error for
column (4) can be recovered by dividing column (4) by column (7). We
give the S/N as opposed to the error since the S/N was used to
define the catalog cutoff and later is used in the calculation of
log N- log S. In column (8), we give the best fit deconvolved size in
arcseconds. If a resolved two dimensional
Gaussian was the best fit, we give the major and minor axis size
(FWHM) and the position angle. Upper limits are given for sources
which were unresolved based on the results of JMFIT or SAD.  When the
minor axis is unresolved, we give ``0'' as the minor
axis size in the table and assume a one-dimensional Gaussian size when
estimating the total flux density. 
sources with very large sizes, for which only a largest angular
size is given in column (8), sizes and total flux densities
were estimated directly
from the images using the AIPS routines, IMVAL and TVSTAT.

\subsection{Angular size distribution}

\begin{figure}
\plotone{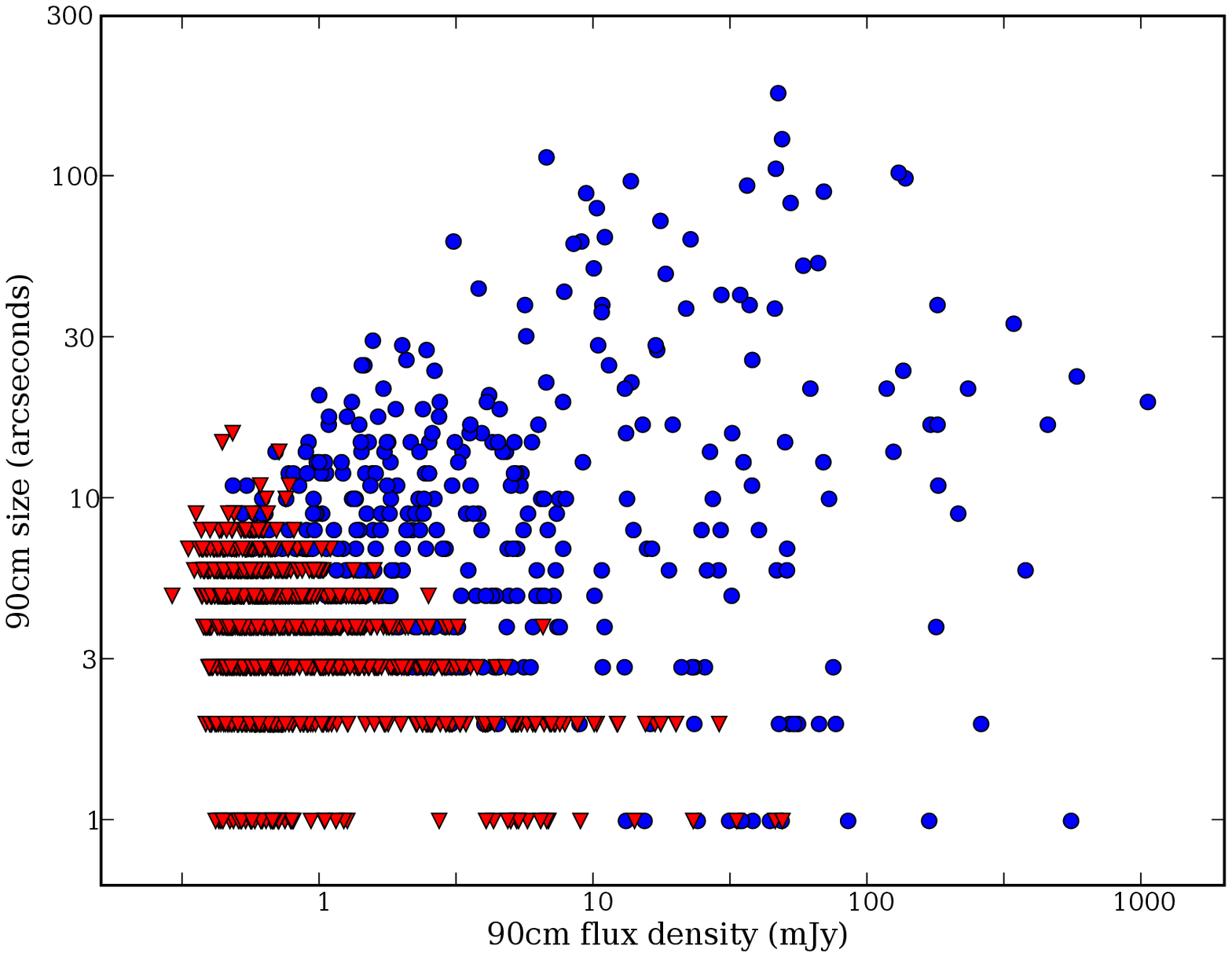}
\caption{Log-Log plot of angular size versus 90cm flux density using
the data in Table~\ref{S90}. Blue dots are for significantly resolved
sources while the red triangles represent sources with upper limits
to the angular size. 
\label{af90}}
\end{figure}

\begin{figure}
\plotone{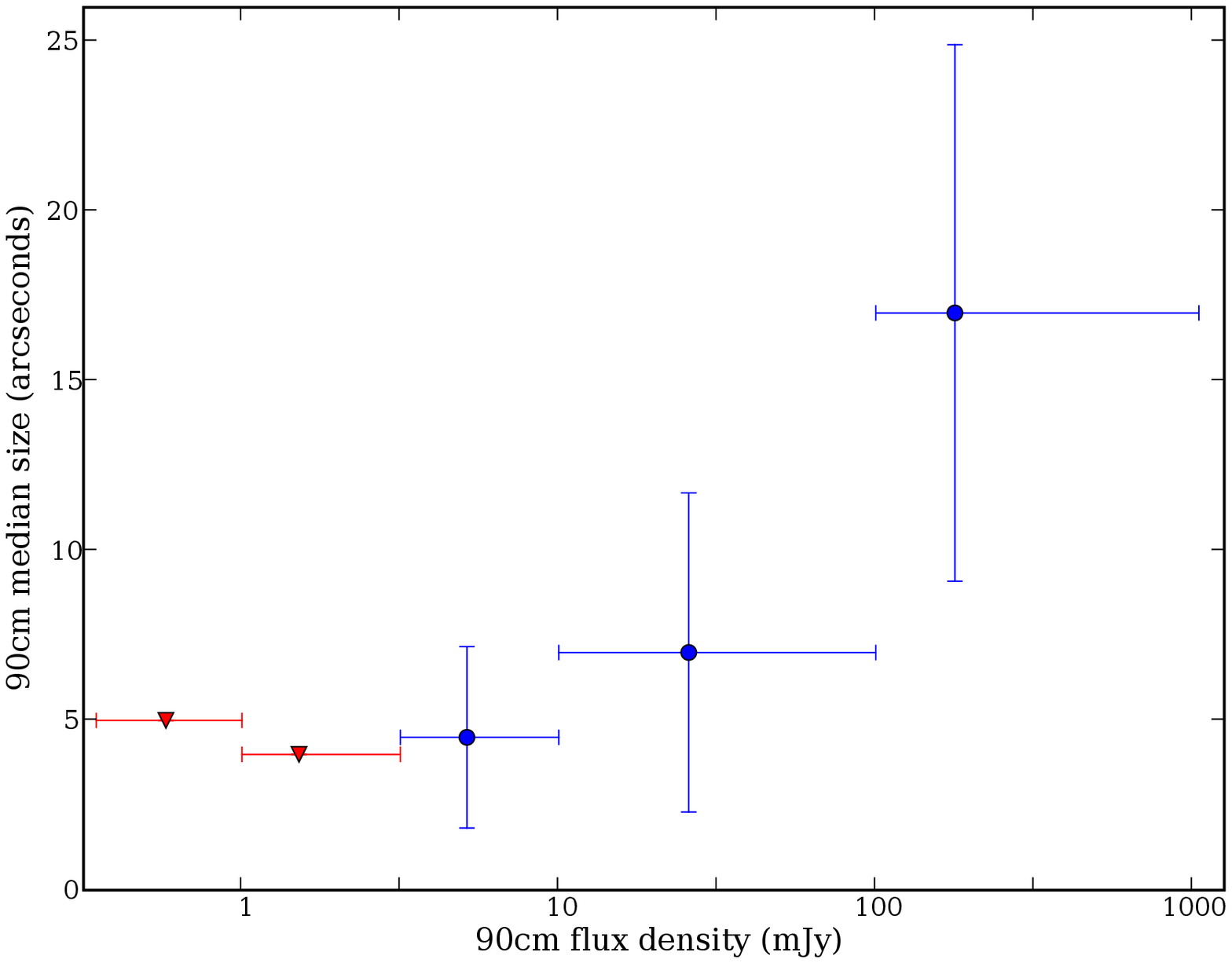}
\caption{Observed median angular size for various different 
logarithmic ranges in 90cm flux density. Blue dots show significant
detections of the median while the red triangles are upper limits to the
median. The error bars in the flux densities show the size of the
bins used to calculate the statistics, while the error bars in the 
sizes are estimated errors in the median values. \label{mf90}}
\end{figure}

The 90cm catalog covers a larger area than the 20cm survey but at
a significantly lower sensitivity. This means that the 90cm survey
samples sources in a higher
range in 20cm flux density with different properties in angular size
and absolute radio luminosity than our 20cm survey.  In Figure~\ref{af90} we show the median
angular size -- 90cm flux density distribution. Above 3 mJy we resolve most of
the sources
but below 3mJy most of the sources are relatively small. Our 20cm
results
for the same sources reported in paper I show that the typical median
size is $\sim 1.2$\arcsec\ .  This trend in
the measured  median sizes at 90cm is shown in Table~\ref{sss} and
Figure~\ref{mf90}. 
Although similar resolution data have not been reported at 90cm these
results are also consistent with previous results at higher
frequencies, extrapolated to 90cm 
\citep{w03}.

Some of the 90cm sources have larger sizes than the corresponding
20cm counterparts. For these sources we checked the flux
densities and sizes at 20cm by smoothing to the resolution of the 90cm detection
images. For most of these sources the lower resolution 20cm sizes and
flux densities agree well with the fits at the original resolution. 
The total flux density changed by more than 10\% 
for only three individually cataloged 20cm sources (00061, 01186
and 01193). In making the 20cm catalog, the larger flux densities at lower resolution
were rejected because the S/N was significantly higher on 
higher resolution images which fitted smaller sizes. The 90cm survey
is also more sensitive to smaller
spatial frequencies and any steeper spectrum, more extended emission. 
 While these results do not change any of the conclusions in
paper I, we are likely a little 
incomplete for the largest sources at 20cm, especially near the bottom of the catalog. 
For the spectral index analysis below we have used the 20cm fits from
imaging at 90cm resolution.

\subsection{Spectral Indices}

\subsubsection{90cm selected sources}

For our spectral index analysis we restrict ourselves to sources
within 20\arcmin\ of the field center. At that radius 
we are complete at 20cm and beyond that distance
from the field center the uncertainty due to the primary beam
correction at 20cm becomes significant. 
Since the 20cm image is so deep, almost all the 90cm
sources within 20\arcmin\ of the field center have counterparts which we
can use to determine spectral indices. Two 90cm sources (01492 and 01501)
were detected on the 20cm 6\arcsec\ resolution image only with $S/N < 5$.
Three more sources (00766 $\alpha>1.46$, 01346 $\alpha>1.71$ 
and 01491 $\alpha>1.84$) have no 20cm counterpart with a $S/N > 3$.
However, these three sources all have low 90cm S/N detections, $5.0 < S/N < 5.6$,
and have no counterparts
on our deep optical/NIR images \citep{p09} or our deep 50cm GMRT image
\citep{o09}. These sources  either have very steep spectra and are very distant
or they are spurious. We choose to leave them out of the
following discussion but note that there may be a small, very steep
spectrum tail to the distribution.

In most other cases, the positions and sizes of individual sources
in 20cm and 90cm  catalogs agree well with one another and the spectral index
is calculated from the values listed in the respective tables. In cases where
the cataloged 20cm and 90cm parameters do not agree well in size
and/or position the sources were investigated on the respective 20cm
and 90cm images by eye. When more than one 20cm source corresponds to
a 90cm source the flux densities of the 20cm sources were summed.
If only the 90cm source is apparently resolved but the 20cm source
size implies that the 90cm source should be unresolved, then the
peak flux on the highest resolution 90cm image and the total
20cm flux density were used to calculate the spectral index (e.g. 90cm
source 00727). If both the 20cm and 90cm source are resolved then the peak
flux density at both 20cm and 90cm is used from images at
the resolution of the cataloged 90cm source (e.g. 90cm source 12407).
In Table~\ref{Si90} we give the
first ten lines of the electronic table summarizing the 90cm-selected
spectral indices, including 90cm source 00727 and 12407 discussed
above. If we needed to sum more than one 20cm flux density, then we
give the total in column 5. Otherwise we give the cataloged total
flux densities but indicate with a note the few cases where 
discrepant size estimates drive us to use peak flux densities for
the corresponding spectral index estimate. 

\begin{figure}
\plotone{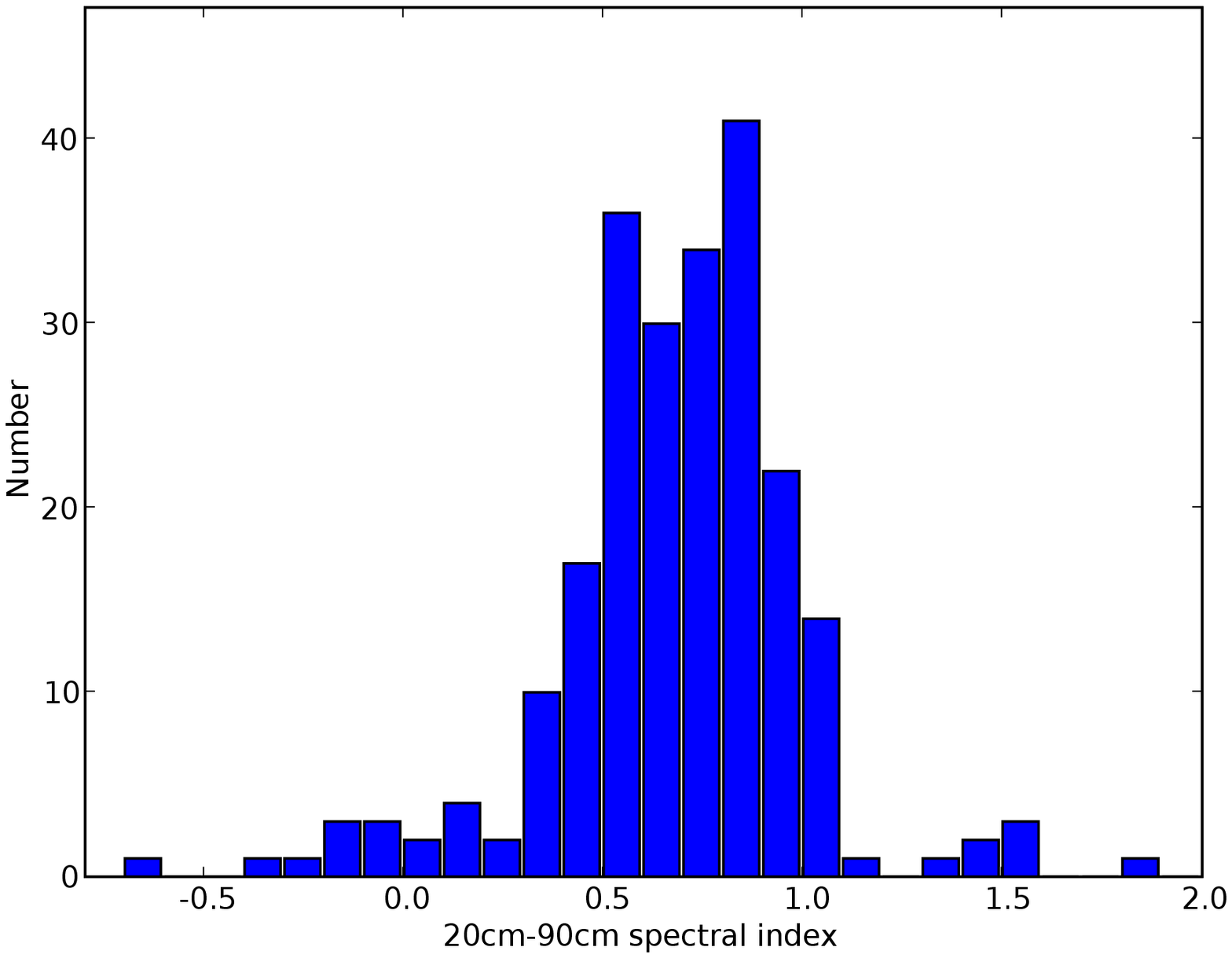}
\caption{Histogram of 90cm to 20cm spectral indices for sources 
detected above 5 sigma at 90cm in the region cataloged for the
20cm survey. \label{sp90}} 
\end{figure}

Figure~\ref{sp90} contains the observed histogram of 90 to 20cm
spectral indices ($S\propto \nu^{-\alpha}$) for sources detected
above 5 sigma at 90cm.
In Table~\ref{sis90} we summarize the statistics for these 
sources as a function of 90cm flux density and size. The mean and
median spectral indices are $\sim 0.7$ for all of our subsets.   
This value is 
flatter than the mean 408-1407 MHz spectral index of 0.92
from 5C12 for sources detected at 408 MHz above
40 mJy \citep{Be82}. We have only a few sources with 90cm flux
densities
$> 40$ mJy in our 90cm survey, so our results are consistent with a flattening in the
mean spectral index below the 5C12 characteristic flux density. 
The deeper LBDS 327 MHz survey has a 90cm-selected,
327-1462 MHz median spectral index shifting from $\sim 0.9$ above 100 mJy,
to $\sim0.7$ between 10 and 100 mJy, and then down to $\sim0.5$ between 3.6 and
10 mJy \citep{Oort}. We agree with their estimate in the $10-100$ mJy
range but do not find the flatter median spectral index below 10 mJy
near the bottom of the LBDS survey. 

\subsubsection{20cm selected sources}

Most recent work on deep fields involves surveys at 20cm, so
it is also interesting to study the spectral index distribution
when selecting the sources at 20cm. Since
most of the 20cm sources are not detected
directly at 90cm, we must use a combination of a high 20cm flux
density cutoff to study the stronger 20cm sources individually and stacking subsets at
90cm in order to study the fainter 20cm population.
Most of the weaker sources are much smaller than the size of the
synthesized beam of our 90cm survey (see paper I).  Those sources that
are expected to be unresolved at 90cm are suitable for stacking
analysis.

\begin{figure}
\plotone{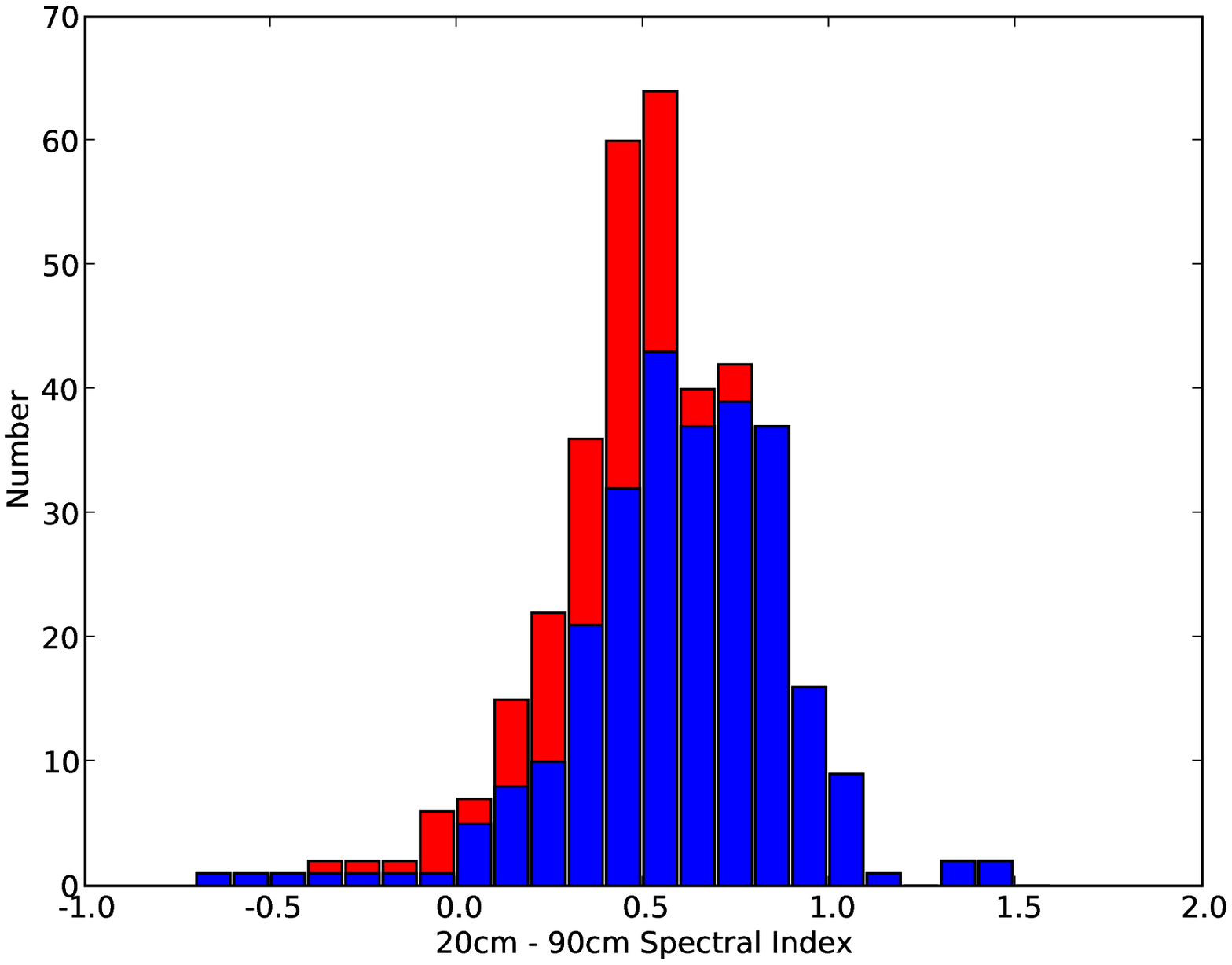}
\caption{Histogram of 90cm to 20cm spectral indices for sources with
20cm flux densities $> 100\mu$Jy. Blue: measured spectral indices for
90cm sources with S/N $ >3$, Red: 3 sigma upper
limits for 90cm sources with  S/N $<3$. \label{sp20}} 
\end{figure}

\begin{figure}
\plotone{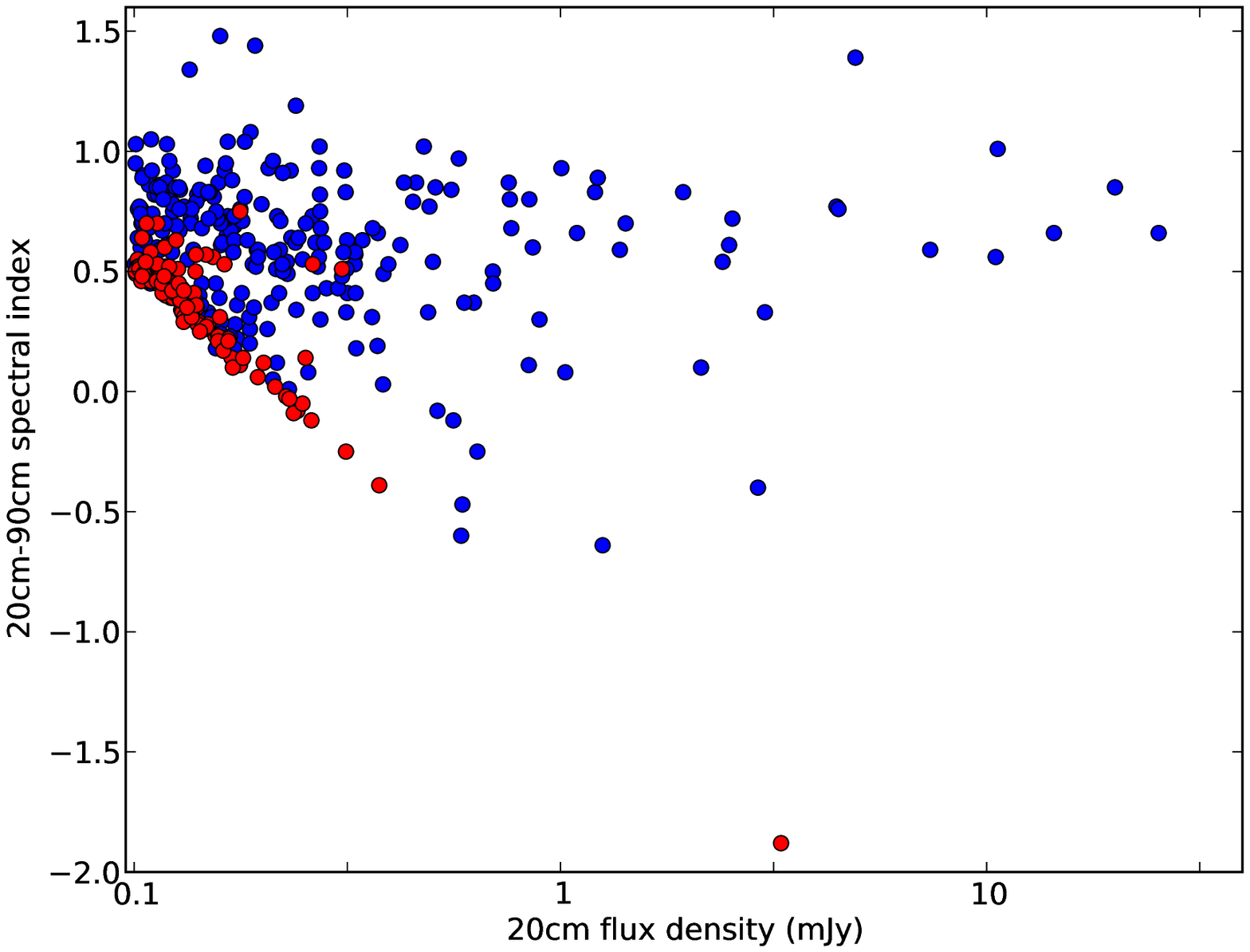}
\caption{90cm to 20cm spectral indices versus 20cm flux density
plotted on a logarithmic scale.
Blue dots are sources with 90cm detections $>3$ sigma. Red dots are
sources with 3 sigma 90cm upper limits.\label{sp20f}} 
\end{figure}

For the high flux density subset, we select only 20cm sources with
total flux densities $> 100 \mu$Jy from paper I within 20\arcmin\ of
the field center.  In Table~\ref{Si20} we show the first 10 lines
of the electronic table of the 90cm--20cm spectral indices for
these sources.  In Figure~\ref{sp20} we plot the
histogram of the spectral indices of these sources against their 20cm
flux densities.  Since, in this subset of sources, many of the
sources are not detected above 3 sigma at 90cm, we plot the blue
area for sources detected above 3 sigma while the red area
represents the 3 sigma upper limit corrected for the observed 20cm  
source size.  The median spectral index is 0.52(0.04) including
the upper limits, significantly flatter than the 90cm selected sample
and samples with brighter limiting flux densities selected at 20cm.  In
Figure~\ref{sp20f} we show spectral index plotted against the
20cm flux density for sources with $S_{20} > 100\mu$Jy.  This plot 
shows the sources with 90cm upper limits as red circles clearly
delineating the section of the plot which is not allowed due to
the 90cm sensitivity.  Even with this high limit on the 20cm catalog,
37\% of the sources with $100\mu$Jy$ < S_{20} < 1000\mu$Jy have upper
limits at 90cm; thus we cannot determine the details of the dependence
of the spectral index distribution on 20cm flux density for $S_{20}<
1$ mJy from the properties of individual sources.

In order to study the spectral index distribution down to the bottom
of the 20cm catalog, we need to use stacking on the 90cm image.
Since we need to restrict the stacks to angular sizes which are
effectively unresolved, we omit a modest, but significant number
of sources that are resolved by the 90cm synthesized beam.
In the following discussion, we divide the sources into subsets by
ranges of 20cm flux density and also split the subsets into sources
which are $>3$\arcsec\ and $\le 3$\arcsec.  In Paper I we found
that sources with $S_{20}< 1$ mJy have a median size of $\sim
1$\arcsec, unlike the stronger sources individually detected at 90cm.
If we omit from the stacking analysis the $\sim 10$\% of
sources with $S_{20}< 1$ mJy that have sizes $> 3$\arcsec\ , we
expect that the spectral properties of the $\le3$\arcsec\
subsets should be close to those of the full population.
In stacking each subsample with flux densities at 20cm $<1$mJy,
we extract the observed 90cm brightness in $\mu$Jy/beam at the
position of the 20cm source.  For small sources this should be a good
estimate of the total flux density in $\mu$Jy.  Within the subset,
we then calculate the mean and median flux densities at both
frequencies.  The spectral index computed from the two mean flux
densities and that computed from the two median flux densities are
listed for each subset in the bottom six lines of Table~\ref{sis20}.
The errors in Table~\ref{sis20} are calculated assuming counting
statistics from the total number in each subset, combined in
quadrature with an assumed standard deviation for the population of
spectral indices of 0.30 which seems appropriate based on the higher flux
density, 90cm-selected, spectral indices but is only an educated
guess.

\begin{figure}
\plotone{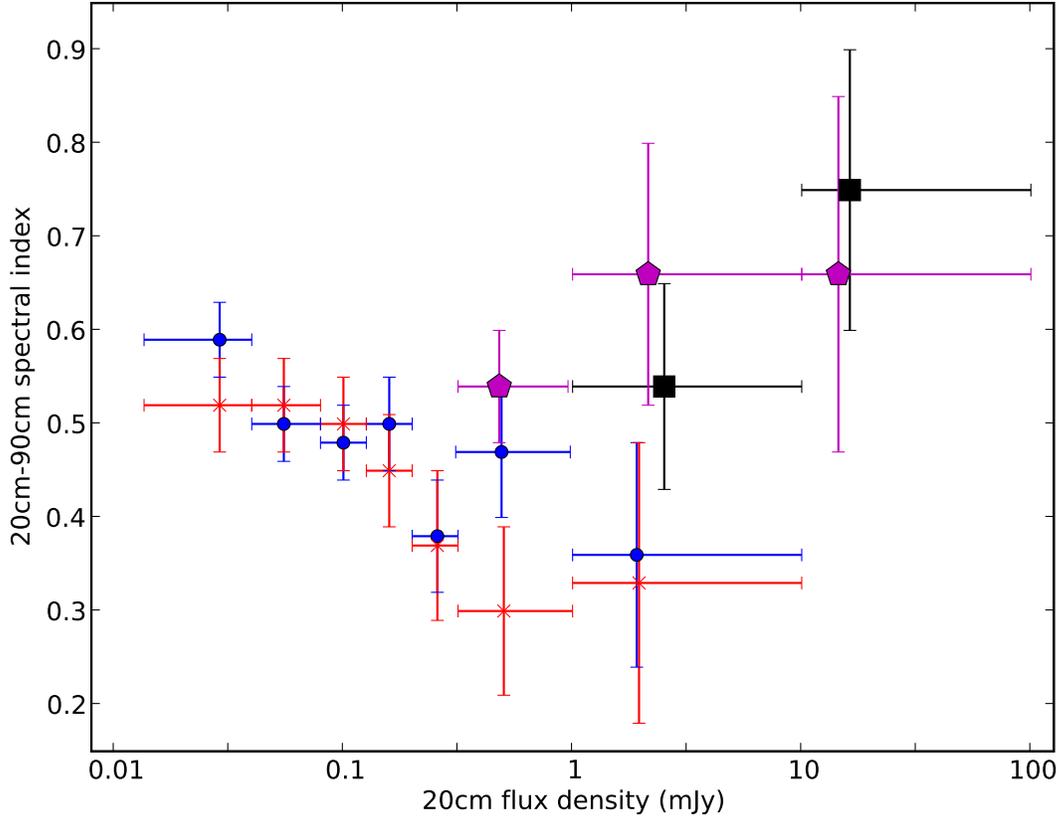}
\caption{20cm selected spectral index versus 
20cm flux density in mJy in various intervals in 20cm flux density: 
The results are from table~\ref{sis20}. The black squares and the
magenta pentagons are the means and medians respectively for the entire sample.
The blue circles and red crosses are the means and medians respectively
for sources with 20cm
sizes $\le3$\arcsec\  using stacking of 90cm cutouts as described in
the text. The flux density error bars show the sizes of the bins while
the spectral index error bars show the estimated error in mean or
median. \label{siall}}
\end{figure}

For sources stronger than 1 mJy at 20cm, all except one have a
detection at 90cm.  Excluding this source we can calculate the median
and mean spectral index for subsets selected by source flux density
and size and these are listed in  Table~\ref{sis20}. In
Figure~\ref{siall} we summarize the results graphically.
Above 1 mJy, where we can study both size subsets, the more
resolved sources have steeper spectra than the $\le 3$\arcsec\ subset.
The mean and median spectral indices continue to flatten for the full
population.  For the small sources, the median and mean spectral
indices flatten to $\sim 0.3-0.5$.  Below 1 mJy the small source
medians and means continue to be in the range $\sim 0.3-0.5$, but
with a clear trend to steeper spectra at the lowest flux densities in
the 20cm sample.

\subsubsection{90cm versus 20cm selection}

The results for the spectral indices are quite
different depending on whether we select the samples at 90cm or 20cm.
At 90cm we find very constant mean and median spectral index of 
$\sim 0.7$ down to the
survey limit of $300\mu$Jy. No obvious dependence is
found on angular size. Approximately the same result is found 
in the 90cm selected sample down
to 1 mJy at 20cm which would correspond to $\sim 3$ mJy at 90cm with
a spectral index $\sim 0.7$. However, sources
$> 3$\arcsec\ in size tend to have steeper spectra than those $\le 3$\arcsec.
Below 1 mJy at 20cm, many more sources do not have detections than would be 
expected if the $\sim 0.7$ spectral index continued. For these and
weaker sources, we
are forced to consider only small sources since the surface brightness
sensitivity at 90cm is not high enough to detect sources of all angular sizes.
For this population of small sources, we find much flatter median and mean spectral
indices, $\sim 0.3-0.5$. This result suggests that there is a smaller size,
20cm source population which is attenuated enough at 90cm, relative to
an $\alpha_{20}^{90} \sim 0.7$, to affect the median source properties
relative to a 90cm-selected population. 

\subsection{Log N - Log S}

The calculation of the 90cm log N - log S is much easier than for 20cm
in paper I because 1) we only consider a radius of one degree which doesn't
reach the half power point of the primary beam and 2) the synthesized beam
is bigger, $\sim 6.2$\arcsec\ so fewer sources are resolved, and 3) 
the fractional channel bandwidth is
smaller. The these  parameters dramatically reduce the problems we faced in paper I
and allow us to perform a simpler analysis. Only below 1 mJy is the
incompleteness due to source size an issue. For  weak sources we use the
same formalism as in paper I to account for missing sources due to resolution and bandwidth
smearing and the same assumed source size distribution. 
Even for the weaker sources, since the resolution is lower, 
the impact of these corrections is very small. 
In Table~\ref{counts}, we summarize our results for the 5 sigma
catalog only, 
unlike paper I where we performed a more complicated calculation
with a variable $S/N$ cutoff.

\begin{figure}
\plotone{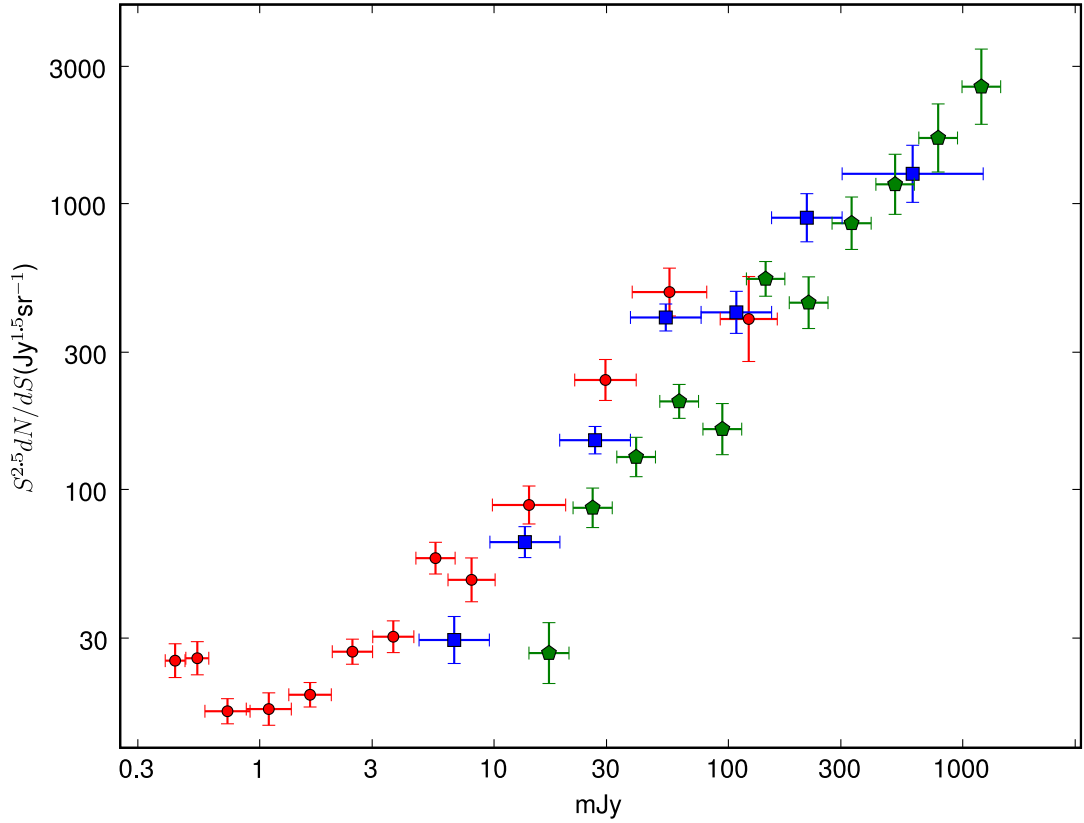}
\caption{90cm log N -- log S differential distribution. Green pentagons are
from the 5C12 survey at 408 MHz scaled to 324.5 MHz using a mean
spectral index of 0.9 \citep{Be82}. Blue squares are from 327 MHz observations
with Westerbork of the LBDS \citep{Oort}. Red circles are from  
table~\ref{counts} in this paper. \label{logNS}}
\end{figure}

In Figure~\ref{logNS}, we plot our results along with the 327 MHz results
for LBDS \citep{Oort} for their 5 sigma catalog and the 5C12 results from
408 MHz scaled by their quoted mean spectral index of 0.9 to 324.5 MHz
\citep{Be82}. The results are in general agreement and our additions show that
we have reached the flat region of the counts seen at other frequencies near
3 mJy. This change in slope also corresponds roughly to the minimum in the
spectral index distribution seen in Figure~\ref{siall}.

\section{Discussion}

These data fill out the picture of the meter-wavelength radio sky
a bit more clearly, building on the earlier work cited above,
but also raise new questions.  Instead of finding a large steep
spectrum population, we find flatter spectra for the subset of
sources with angular sizes  $< 3$\arcsec and $S_{20}< 10$mJy.  We also find
that the differential log N - Log S at 90cm flattens in roughly
the same flux density range ($S_{90}< 5$ mJy).  Thus the nature of
the meter wavelength population seems to be changing in the few mJy
range.  Often a trend toward flattening radio spectra at higher
frequencies and the corresponding change in slope of log N - log S
is attributed to star-forming galaxies becoming dominant
\citep[e.g.,][]{w03}.  However, only for 20cm luminosities $<10^{23}$
W Hz$^{-1}$ are star-forming galaxies more common than AGN
\citep[e.g.,][]{c02}. 
For sources with $S_{20}> 1$mJy (equivalent to $S_{90} >
2$mJy with $\alpha_{20}^{90} \sim 0.5$) to have a 20cm 
luminosity $ < 10^{23}$ W Hz$^{-1}$, their redshift would have to
be $< 0.2$.  This redshift is much too
low for most such mJy sources to be dominated by
star-formation, since we find a median $z\sim 1$ for our sample \citep{p09}.
 Thus it seems likely that the 
changes we are observing are in the AGN population. Furthermore, 
spectral studies at wavelengths $< 20$cm, show that the typical
$\mu$Jy source has a quite steep spectrum, $\sim 0.8-0.9$
\citep{f06}.  Therefore our 20cm-selected $\mu$Jy sources with flatter
spectra cannot be due to free-free emission, since the flat spectrum
free-free emission should be less important at longer wavelengths.

One might think that flatter spectra in AGN might be due to
synchrotron self-absorption as is seen in many beamed radio galaxies
and quasars. However, our work in paper I and other studies 
\citep[e.g.,][]{m05,f06} show that the typical sizes for these sources
are $\sim$1\arcsec. This argues against synchrotron self-absorption
being dominant since that mechanism requires sizes $\ll1$
milliarcsecond to be important. One
possibility is that 
the flatter spectra could result from the combination of 
a relatively flat-spectrum AGN jet with a spectral index $\sim 0.5$ 
\citep[e.g.,][]{bp,bwk} and a synchrotron self-absorbed core.

Since the sources are relatively small, free-free absorption is
a possibility. Free-free
absorption for such sources depends in general on the details of the
clumpiness of the absorbing thermal gas and its geometric relation
to the synchrotron emitting medium. In star-forming systems which have
been well studied, the radio emission is extended more uniformly throughout
the galaxy than the dust or the HII regions \citep{h98,m08}. Free-free
absorption is also seen on small scales in some radio AGN
\citep[e.g.,][]{w94,g04}. For a uniform
density foreground medium with a temperature $\sim 10^4$K, the characteristic
free-free absorption turnover frequency is 
$\nu_t$(MHz)$\sim 0.5n$(cm$^{-3}$)$l^{0.5}$(pc), where $l$ is the
projected pathlength.
For example, for a typical source in
our sample with a size of 1\arcsec at $z\sim 1$, a typical radius is
$\sim 4$ kpc and, for a rest frame turnover frequency of $\sim 300$ MHz, one
needs a density of 10 cm$^{-3}$, 
which would correspond to a mass of $\sim 6\times 10^{10}$
$M_{\odot}$ if uniformly distributed in a sphere. A more realistic model with
a smaller filling factor and a different geometry  reduces
the mass estimates but if free-free absorption is important then a reasonably
large mass of ionized gas must be involved.  The flattening could also
be due to ionization
losses as has been suggested for some star-forming galaxies 
\citep[e.g.,][]{t06}.
In any case, the flatter spectra and the flatter counts observed below 
3 mJy suggest a
change in the nature of the population below this flux density level
which is not well understood.  We will discuss this point further
in future papers where we add redshifts and data from other
wavelengths to
the analysis.

\section{Conclusion}

We have presented a very sensitive 90cm image of the SWIRE
deep field. This image combined with our uniquely sensitive 20cm
image of this field allows us to study the meter wavelength 
spectral indices as a function of flux density for the $\mu$Jy
radio population. For 90cm-selected sources the properties of
the sources are consistent with previous work with a mean spectral
index near 0.7 and few very steep spectrum sources. Thus
no large population of very steep spectrum $\mu$Jy sources seems
to exist down to our limiting flux density.

For the subset of sources selected at 20cm with
sizes $<3$\arcsec\ 
(which contains about $\sim 90$\% of all 20cm-selected sources $< 1$
mJy), the mean and median spectral indices flatten from 
$\alpha \sim 0.7$ to $\alpha \sim 0.3-0.5$ below 10 mJy with a
 trend toward steeper spectra at the lowest flux densities.
The 90cm log N-log S counts flatten below 5 mJy as they do at corresponding
low flux densities at higher frequencies. The change in the source
properties at a few mJy is not well understood but probably involves
the AGN population, not primarily star-formation-dominated galaxies.

\clearpage

\begin{deluxetable}{rrrr}
\tablecolumns{4}
\tablewidth{0pt}
\tablecaption{Observing Runs Summary\label{OR}}
\tablenum{1}
\pagestyle{empty}
\tablehead{
\colhead{Configuration} & 
\colhead{Start date} &
\colhead{End date} &
\colhead{Hours}}
\startdata	
A&06Feb19&06May17&77.1\\
C&07Jan04&07Jan05&7.9\\
\enddata
\end{deluxetable}
\clearpage
\begin{deluxetable}{lrrrrrrl}
\tablecolumns{7}
\tablewidth{0pt}
\tablecaption{Radio Source Catalog\label{S90}}
\tablenum{2}
\pagestyle{empty}
\tablehead{
\colhead{Name} & 
\colhead{RA(2000.0)} &
\colhead{Dec(2000.0)} &
\colhead{Peak} &
\colhead{Total} &
\colhead{Error} &
\colhead{S/N} &
\colhead{Size}\\
\colhead{} &
\colhead{} &
\colhead{} &
\colhead{$\mu$Jy/b} &
\colhead{$\mu$Jy} &
\colhead{$\mu$Jy} &
\colhead{} &
\colhead{\arcsec\ x\arcsec\ pa$=$\arcdeg}}  
\startdata
00018&10 38 14.59(0.03)&59 02 36.0(0.1)&  2778&    5425&  320&   26.7&12x2pa= 91\\      
12017&10 38 24.64(0.10)&59 00 35.1(0.6)&  1423&    2498&  340&   10.9&15x6pa=64\\   
00036&10 38 25.64(0.11)&59 05 58.0(0.5)&   620&     892&  241&    5.5&8x0pa=113\\    
00043&10 38 29.75(0.03)&58 52 05.0(0.2)&  1071&    1071&  115&    9.7&$<$3\\     
12045&10 38 31.50(0.03)&58 47 09.0(0.4)&  3024&    4763&  348&   22.2&14x3pa=13\\    
00051&10 38 34.56(0.07)&59 15 15.3(0.5)&   629&     629&  116&    5.5&$<$6\\     
00058&10 38 37.82(0.01)&59 09 45.2(0.1)&  5423&    5423&  196&   49.8&$<$2\\     
00059&10 38 39.16(0.01)&58 59 12.3(0.1)& 28469&   28570&  874&  271.1&$<$2\\         
00065&10 38 42.27(0.05)&59 01 10.5(0.3)&   919&     919&  118&    8.0&$<$4\\    
00068&10 38 43.23(0.07)&58 43 14.2(0.6)&   594&     594&  117&    5.1&$<$4\\
\enddata
\end{deluxetable}
\clearpage

\begin{deluxetable}{rrrr}
\tablecolumns{4}
\tablewidth{0pt}
\tablecaption{90cm Source Size Statistics\label{sss}}
\tablenum{3}
\pagestyle{empty}
\tablehead{
\colhead{log($S_{90}$)\tablenotemark{a}}& 
\colhead{Num\tablenotemark{b}}&
\colhead{Size\tablenotemark{c}}&
\colhead{log($S_{90}$\tablenotemark{d})}\\
\colhead{}&
\colhead{}&
\colhead{\arcsec}&
\colhead{}}
\startdata	
$5.0-6.1$&20&17.0(7.9)&5.25\\
$4.0-5.0$&95&7.0(4.5)&4.41\\
$3.5-4.0$&126&4.5(2.7)&3.71\\
$3.0-3.5$&335&$<4$&3.18\\
$2.5-3.0$&859&$<5$&2.76\\
\enddata
\tablenotetext{a}{ Range of the log of 90cm flux densities with $S_{90}$ in $\mu$Jy}
\tablenotetext{b}{ Number of sources in interval}
\tablenotetext{c}{ Median size with error in median in parenthesis}
\tablenotetext{d}{ Median of the 90cm flux density of the subsamble in $\mu$Jy }
\end{deluxetable}
\clearpage

\begin{deluxetable}{rrrrrrrrrr}
\tablecolumns{10}
\tablewidth{0pt}
\tablecaption{Spectral Indices for 90cm Selected Sources \label{Si90}}
\tablenum{4}
\pagestyle{empty}
\tablehead{
\colhead{Name} & 
\colhead{$\alpha_{20}^{90}$\tablenotemark{a}} &
\colhead{Error} &
\colhead{$S_{90}$} &
\colhead{$S_{20}$} &
\colhead{Res\tablenotemark{b}} &
\colhead{Size} &
\colhead{Res\tablenotemark{b}} &
\colhead{Size} &
\colhead{Note\tablenotemark{c}}\\
\colhead{90cm}&
\colhead{} &
\colhead{$\alpha_{20}^{90}$} &
\colhead{$\mu$Jy} &
\colhead{$\mu$Jy} &
\colhead{90cm} &
\colhead{\arcsec} &
\colhead{20cm} &
\colhead{\arcsec} &
\colhead{}}  
\startdata
00725&0.57&0.07&758&330.2&$<$&4&r&2.6&\omit\\ 
00727&0.48&0.11&813&211.4&r&7&$<$&2.4&*\\ 
00728&-0.40&0.04&1616&2906.2&$<$&3&r&1.5&\omit\\ 
00733&0.29&0.04&1219&803.6&$<$&1&r&0.8&\omit\\ 
00737&0.48&0.04&3000&1487.5&r&2&r&2.0&\omit\\ 
12407&0.90&0.06&1211&272.4&r&12&r&7.0&*\\ 
00744&0.72&0.12&390&136.3&$<$&6&r&2.8&\omit\\ 
00755&0.46&0.06&2090&1069.1&r&9&r&8.3&\omit\\ 
00756&0.86&0.12&402&114.4&$<$&5&$<$&2.4&\omit\\ 
00761&0.33&0.09&512&314.3&$<$&3&r&1.3&\omit\\ 
\enddata
\tablenotetext{a}{ 90cm-20cm spectral index, $S\propto \nu^{-\alpha}$}
\tablenotetext{b}{ `$<$': next column upper limit, `r': next column deconvolved
major axis }
\tablenotetext{c}{ $*$: Spectral index determined from peak of equal
  resolution images at 20cm and 90cm}
\end{deluxetable}
\clearpage

\begin{deluxetable}{lrrrrrr}
\tablecolumns{7}
\tablewidth{0pt}
\tablecaption{Spectral Index Summary for 90cm Selected Sources\label{sis90}}
\tablenum{5}
\pagestyle{empty}
\tablehead{
\colhead{log($S_{90}$)\tablenotemark{a}}& 
\colhead{Num\tablenotemark{b}}&
\colhead{$\alpha$\tablenotemark{c}}&
\colhead{$\alpha$\tablenotemark{c}}&
\colhead{std\tablenotemark{d}}&
\colhead{log($S_{90}$\tablenotemark{e})}&
\colhead{log($S_{90}$\tablenotemark{e})}\\
\colhead{}&
\colhead{}&
\colhead{mean}&
\colhead{med}&
\colhead{dev}&
\colhead{mean\tablenotemark{f}}&
\colhead{med\tablenotemark{f}}}
\startdata
All\\
$2.5-5.0$&229&0.68(0.02)&0.70(0.03)&0.31&3.38&2.77\\
$4.0-5.0$&11&0.70(0.05)&0.66(0.07)&0.17&4.48&4.37\\
$3.0-4.0$&56&0.68(0.04)&0.76(0.06)&0.34&3.37&3.23\\
$2.5-3.0$&162&0.68(0.02)&0.70(0.03)&0.30&2.72&2.67\\
$>3$\arcsec\\
$2.5-5.0$&55&0.71(0.03)&0.72(0.04)&0.21&3.68&3.02\\
$4.0-5.0$&7&0.75(0.08)&0.77(0.10)&0.19&4.43&4.24\\
$3.0-4.0$&22&0.74(0.05)&0.76(0.07)&0.24&3.40&3.25\\
$2.5-3.0$&26&0.68(0.04)&0.67(0.05)&0.19&2.77&2.76\\
$\le 3$\arcsec\\
$2.5-5.0$&174&0.67(0.03)&0.70(0.03)&0.34&3.22&2.70\\
$4.0-5.0$&4&0.60(0.04)&0.61(0.05)&0.06&4.54&4.49\\
$3.0-4.0$&34&0.65(0.07)&0.67(0.08)&0.38&3.37&3.23\\
$2.5-3.0$&136&0.68(0.03)&0.71(0.04)&0.19&2.77&2.77\\
\enddata
\tablenotetext{a}{ Range of the log of 90cm flux densities with
  $S_{90}$ in $\mu$Jy}
\tablenotetext{b}{ Number of Sources in the interval.}
\tablenotetext{c}{ Spectral Index between 90cm and 20cm defined as
 $S \propto \nu^{-\alpha}$ .}
\tablenotetext{d}{ Estimated standard deviation in the population}
\tablenotetext{e}{ $\mu$Jy}
\tablenotetext{f}{ The log of the mean or median of each subsample is given in this column.}
\end{deluxetable}
\clearpage

\begin{deluxetable}{rrr}
\tablecolumns{3}
\tablewidth{0pt}
\tablecaption{Spectral Indices for 20cm Selected Sources \label{Si20}}
\tablenum{6}
\pagestyle{empty}
\tablehead{
\colhead{Name}& 
\colhead{$\alpha_{20}^{90} $\tablenotemark{a}}&
\colhead{Error}\\
\colhead{20cm}&
\colhead{}&
\colhead{$\alpha_{20}^{90}$}}
\startdata
00013&0.57&0.07\\ 
00016&-0.40&0.04\\ 
00021&1.02&0.06\\ 
00024&0.58&0.18\\ 
00028&0.86&0.12\\ 
00029&0.33&0.09\\ 
00030&0.82&0.10\\ 
00033&$<$0.28&\omit\\
00037&$<$-0.12&\omit\\
00044&-0.64&0.10\\ 
\enddata
\tablenotetext{a}{ 90cm-20cm spectral index, $S\propto \nu^{-\alpha}$}
\end{deluxetable}
\clearpage

\begin{deluxetable}{lrrrrr}
\tablecolumns{6}
\tablewidth{0pt}
\tablecaption{20cm Selected Spectral Index Summary \label{sis20}}
\tablenum{7}
\pagestyle{empty}
\tablehead{
\colhead{log($S_{20}$)\tablenotemark{a}}& 
\colhead{Num\tablenotemark{b}}&
\colhead{$\alpha$\tablenotemark{c}}&
\colhead{$\alpha$\tablenotemark{c}}&
\colhead{$log(S_{20}$\tablenotemark{d}$)$}&
\colhead{$log(S_{20}$\tablenotemark{d}$)$}\\
\colhead{}&
\colhead{}&
\colhead{mean}&
\colhead{med}&
\colhead{mean\tablenotemark{e}}&
\colhead{med\tablenotemark{e}}}
\startdata
All\\
$2.5-3.0$&43&\omit&0.54(0.06)&\omit&\omit\\
$3.0-5.0$&24&0.58(0.09)&0.66(0.11)&3.73&3.40\\
$4.0-5.0$&5&0.75(0.15)&0.66(0.19)&4.21&4.16\\
$3.0-4.0$&19&0.54(0.11)&0.66(0.27)&4.22&4.15\\
$>3$\arcsec\\
$3.0-5.0$&12&0.75(0.10)&0.76(0.12)&3.71&3.57\\
$4.0-5.0$&2&0.93(0.30)&0.93(0.36)&4.19&4.19\\
$3.0-4.0$&10&0.71(0.11)&0.74(0.14)&3.49&3.44\\
$\le 3$\arcsec\\
$3.0-5.0$&12&0.42(0.14)&0.60(0.17)&3.75&3.35\\
$4.0-5.0$&3&0.62(0.15)&0.57(0.19)&4.19&4.16\\
$3.0-4.0$&9&0.35(0.18)&0.59(0.22)&3.28&3.29\\
$2.5-3.0$&28&0.47(0.07)&0.30(0.09)&2.69&2.70\\
$2.3-2.5$&35&0.38(0.06)&0.37(0.08)&2.41&2.41\\
$2.1-2.3$&87&0.49(0.05)&0.45(0.06)&2.20&2.20\\
$1.9-2.1$&180&0.48(0.04)&0.50(0.05)&2.00&2.00\\
$1.6-1.9$&485&0.50(0.04)&0.52(0.05)&1.74&1.74\\
$1.0-1.6$&479&0.59(0.04)&0.52(0.05)&1.46&1.46\\
\enddata
\tablenotetext{a}{ Range of the log of 20cm flux densities in the
  interval with $S_{20}$ in $\mu$Jy}
\tablenotetext{b}{ Number of Sources in the interval}
\tablenotetext{c}{ Spectral Index between 90c and 20cm defined as
 $S \propto \nu^{-\alpha}$}
\tablenotetext{d}{ $\mu$Jy}
\tablenotetext{e}{ The log of the mean or median of each subsample is
  given in this column.}
\end{deluxetable}
\clearpage

\begin{deluxetable}{rrr}
\tablecolumns{3}
\tablewidth{0pt}
\tablecaption{90cm differential normalized source counts for 1046+59. The
  table contains 1) $S_l$ (the lower flux density limit of the bin), 
2) $S_h$(the upper flux density limit if the bin), 3) the normalization
  factor and its error. \label{counts}}
\tablenum{8}
\pagestyle{empty}
\tablehead{ 
\colhead{$S_l$}&
\colhead{$S_h$}&
\colhead{$S^{2.5}dN/dS$}\\
\colhead{$\mu$Jy}&
\colhead{$\mu$Jy}&
\colhead{Jy$^{1.5}$sr$^{-1}$}}
\startdata	
375&475&$25.4\pm 3.7$\\
475&600&$25.9\pm 3.7$\\
600&900&$16.9\pm 1.8$\\
900&1350&$17.2\pm 2.4$\\
1350&2000&$19.3\pm 2.0$\\
2000&3000&$27.3\pm 2.9$\\
3000&4500&$30.8\pm 4.2$\\
4500&6750&$58.0\pm 7.9$\\
6750&10000&$48.7\pm 9.4$\\
10000&20000&$89.0\pm 14.7$\\
20000&40000&$243.6\pm 43.7$\\
40000&80000&$495.6\pm 105.1$\\
80000&160000&$398.1\pm 162.5$\\
\enddata
\end{deluxetable}
\clearpage

\end{document}